\documentclass{aa}             
\usepackage{amssymb,graphicx}  
\begin{document}

\title{Truncation of geometrically thin disks around massive black
holes in galactic nuclei}
\author{B.F. Liu \inst{1,2} and E. Meyer-Hofmeister \inst{1}}
\offprints{Emmi Meyer-Hofmeister}
\institute{Max-Planck-Institut f\"ur Astrophysik, Karl
Schwarzschildstr.~1, D-85740 Garching, Germany
\and
Yunnan Observatory, Academia Sinica. P.O.Box 110, Kunming 650011, China
} 

\date{Received:s / Accepted:}

\abstract{
The concept of an advection-dominated accretion flow (ADAF), with or
without wind loss, was used to describe the spectra of the galactic
center source Sgr$\rm{A}^*$, low-luminosity AGN and
nuclei of elliptical galaxies (including M87).
The spectral fits of various authors show that the transition from the
geometrically thin disk to the hot flow occurs at quite different distances,
apparently not uniquely related to the mass flow rate in the disk.
We compare these results with the results of theoretical modeling
where we determine the truncation of the thin outer disk from
the efficiency of mass evaporation. The physics is the same as in the
case of galactic black holes systems (Meyer et al. 2000b). For the
observationally indicated mass flow rates our model predicts a
truncation at $10^3$ to $10^4$ Schwarzschild radii. We discuss whether
far inside this truncation an innermost cool thin disk could exist and
affect the spectrum.
}
\titlerunning {Truncation of geometrically thin disks in AGN}
\maketitle
%

\section{Introduction}
The form of accretion around black holes depends on the accretion
rate. For low, sub-Eddington accretion rates a solution was found in
which the accreting gas has a very low density and is unable to cool
efficiently within an accretion time (Ichimaru 1977, Rees et al. 1982,
Narayan \& Yi 1994, 1995a, 1995b, Abramowicz et al. 1995, Honma 1996,
for a reviews see Narayan et al. 1998, Kato et al. 1998). In this mode
of accretion (advection-dominated accretion flow, ADAF) advection
rather than radiation removes the locally generated accretion heat. 
This is only possible in the inner region. At larger distance the mass
accretes via a cool geometrically thin standard disk.
Esin et al. (1997) considered accretion flows where the ADAF
region extends outward to a certain transition radius, surrounded 
by a cool outside standard disk (with a corona above as
continuation of the hot flow further inward). This picture was
confirmed by the successful modeling of the spectral transitions
observed for Nova Muscae 1991. In this investigation the transition
was taken according to the ``strong ADAF principle'' (Narayan \& Yi
1995).

The transition to the ADAF can be understood as
caused by evaporation of mass from the thin cool disk at the midplane to 
a hot coronal layer above. Such a corona exists predominantly above the
inner region of the cool disk. The investigation of the equilibrium
between these interacting layers shows that a continuous evaporation
of gas occurs. The computation of the vertical structure of the corona
allows to determine the amount of mass which flows inward in the corona at each
distance from the black hole, depending on the mass of the compact star
(Meyer \& Meyer-Hofmeister 1994). A detailed description of this
process with application to disks around black holes is given in Meyer
et al. (2000b). With decreasing distance from the black hole more and
more mass flows in the corona. Where the coronal flow reaches 100\% of
the total accretion flow the cool disk in the midplane ends and
inside is only a hot coronal/advection-dominated flow. The aim of
the present paper is the application of our results to
accretion disks around supermassive black holes.

Coronae in AGN were
considered since many years by several authors (Liang \& Price (1977),
Haardt \& Maraschi (1991), Nakamura \& Osaki (1993), Kusunose \&
Mineshige(1994), Svensson \& Zdziarski (1994), \.Zycki et al. (1995),
Abramowicz et al. (1995) and subsequent workers). For a review on X-ray
spectra of AGN see Mushotzky et al. (1993).

Recently the co-existence of the hot and the cold gas around galactic
black holes and in AGN was investigated in detail by
R\'o\.za\'nska \& Czerny (2000a, 2000b). Their work follows in
principle the same line as the work of Meyer \& Meyer-Hofmeister
(1994) and Meyer et al. (2000b) on the equilibrium between corona
and thin cool disk around a white dwarf or a black hole, but focuses on the
innermost regions near the black hole and incorporates the physics of a
two-temperature corona.

In new investigations of the low luminosity galactic nuclei of M81 and
NGC 4579 (Quataert et al. 1999) evidence was found for thin disks
truncated at about 100 Schwarzschild radii, similar to the result for
NGC 4258 (Gammie et al. 1999). The low luminosity in the elliptical
galaxy M87 (NGC 4486) was pointed out by Reynolds et al. (1996),
illustrating the problem of a ``quiescent'' black hole. The nuclear
regions of six elliptical galaxies were studied by Di Matteo
et al. (1999, 2000). From the X-ray emission together with
recent high-frequency radio observations new constraints for the physical
properties of any proposed ADAF were derived. These radio and
submillimeter observations seem to demand that, at least in some
galaxies, a significant fraction of mass, angular momentum and 
energy is removed from the accretion flow by a wind.

In this paper we first give a short description of the physics of
evaporation, as included in our computations (Sect. 2). In Sect. 3 we
present the results, and compare in Sect. 4 with findings from the
spectra of low luminosity AGN and elliptical galaxies. In Sect. 5 we discuss a
new suggestion for an innermost thin disk which comes from the
evaporation/condensation picture and might help to solve discrepancies
which were already found applying the ADAF model to low-luminosity galaxies.

\section {The equilibrium between thin disk below and corona above: evaporation
of cool gas} 

The equilibrium between a cool disk and the corona above it is investigated
in order to determine how much of the gas flows in the corona, that is,
the amount evaporated from the cool disk into the corona.
This evaporation efficiency strongly depends on the mass of the
compact central object. For a given mass the efficiency increases with
decreasing distance to the center (except in an innermost part).The
co-existence of a cool disk and a corona had already been investigated
for disks in dwarf nova systems (Meyer \& Meyer-Hofmeister 1994, Liu
et al. 1995). The situation is the same in disks around neutron
stars, stellar and supermassive black holes. For the properties of
two-temperature, optically thin ADAFs (Narayan et al. 1998)
successfully applied to accretion in stellar black holes and galactic
nuclei, the quantities accretion rate and distance were written in units of
Eddington accretion rate and Schwarzschild radius. 
($\dot M_{\rm {Edd}}=40{\pi}GM/{\kappa}c$,
$r_{\rm S}=2GM/c^2$, with $M$ mass of central black hole and 
$\kappa$ electron scattering opacity). The solutions then
are invariant. Our relation between evaporation rate and distance is
also invariant.

We solve a set of four ordinary differential equations for 
motion of the gas, mass, energy and thermal flux
(Meyer et al. 2000b) to model the equilibrium between cool disk and
corona near the inner edge of the thin disk. The boundary conditions
are:  (1) at infinity no
pressure (i.e. no artificial confinement, which requires sound
transition at some height) and no influx of heat and (2) at the bottom
of the corona chromosphere temperature and no heat inflow.

\section {Results of computations}

We carried out computations for massive black holes of $2.5\times10^6$
and $10^8 M_\odot$ with Schwarzschild radii of $7.38\times 10^{11}$
and $2.95\times10^{13}$ cm, respectively. In Table 1 and 2 we list
mass flow rates $\dot M$ in the corona for a number of distances $r$.
$\dot M$ and $r$ are also given in units of Eddington accretion rate and
Schwarzschild radius. The relation between $\dot M$ and $r$
is invariant if accretion rates and distances are measured in these
units. The same is true for the coronal temperature (compare Fig. 3
and the comments in Meyer et al. 2000b) and for the fraction of matter
carried away by the wind. Also given are the values of pressure and
mass flow rate into the corona at the lower
boundary. Note that three similar rates are considered: the
evaporation rate is equal to the mass flow rate in the thin disk at
its inner edge; the mass flow rate in the corona is equal to the
evaporation rate minus wind loss (accretion rate onto the black hole,
if no wind loss further in).

\begin{table*}
\caption[]{\label{t:evapr-bh} Evaporation of the disk around 
 a black hole of $2.5\times10^6M_\odot$}
\begin{center}
\begin{tabular}{cccccccc}
\hline
\noalign{\smallskip}
$\log r$&$\log \dot m_0$& $\log P_0$&$ T$(K)&$\dot M=2\pi r^2\dot
m_0 (M_\odot /yr)$&$\lambda$&$\log r/r_{\rm S}$&
$\dot M/{\dot M_{\rm Edd}}$\\
\noalign{\smallskip}\hline\noalign{\smallskip} 
15.87&-11.10&-2.44&$4.51\times10^7$&$4.30\times10^{-5}$&0.23& 4.00 &
$7.81\times10^{-4}$\\
15.48&-9.79&-0.93&$1.10\times10^8$&$1.47\times10^{-4}$&0.20& 3.61 &
$2.67\times10^{-3}$\\
15.00&-8.26&0.88&$3.14\times10^8$&$5.52\times10^{-4}$&0.15& 3.13 &
$1.00\times10^{-2}$\\
14.60&-7.17&2.26&$6.85\times10^8$&$1.09\times10^{-3}$&0.09& 2.73 &
$1.98\times10^{-2}$\\
14.40&-6.72&2.89&$9.59\times10^8$&$1.19\times10^{-3}$&0.04& 2.53 &
$2.16\times10^{-2}$\\
14.30&-6.57&3.14&$1.06\times10^9$&$1.07\times10^{-3}$&0.01& 2.43 &
$1.95\times10^{-2}$\\
14.18&-6.46&3.40&$1.11\times10^9$&$7.77\times10^{-4}$&$\le 10^{-3}$& 2.31 &
$1.41\times10^{-2}$\\

\noalign{\smallskip}\hline\noalign{\smallskip} 
\end{tabular}
\end{center} 
Notation: $\dot m_0$ and $P_0$ vertical mass flow density and pressure at
the lower boundary of the corona, $T$ coronal temperature at the
upper boundary (sound transition), $\dot M$ evaporation rate;
$\lambda$ fraction of mass carried away by
the wind; quantities $r/r_{\rm S}$ and $\dot M/{\dot
 M_{\rm Edd}}$ scaled to Schwarzschild radius
and Eddington accretion rate. $r, \dot m_0, P_0$ in cgs units.
\end{table*}    

\begin{table*}
\caption[] {\label{t:evapr-bh2} Evaporation of the disk around 
 a black hole  of $10^8 M_\odot$}
\begin{center}
\begin{tabular}{cccccccc}
\hline
\noalign{\smallskip}

$\log r$&$\log \dot m_0$& $\log P_0$&$ T$(K)&$\dot M=2\pi r^2\dot
m_0 (M_\odot /yr)$&$\lambda$&$\log r/r_{\rm S}$&
$\dot M/{\dot M_{\rm Edd}}$\\
\noalign{\smallskip}\hline\noalign{\smallskip} 
17.00&-11.12&-2.22&$1.32\times10^8$&$7.49\times10^{-3}$&0.19& 3.53 &
$3.40\times10^{-3}$\\
16.70&-10.16&-1.08&$2.56\times10^8$&$1.74\times10^{-2}$&0.16& 3.23 &
$7.91\times10^{-3}$\\
16.48&-9.49&-0.27&$4.08\times10^8$&$2.90\times10^{-2}$&0.14& 3.01 &
$1.32\times10^{-2}$\\
16.00&-8.32&1.28&$9.63\times10^8$&$4.77\times10^{-2}$&0.04& 2.53 &
$2.16\times10^{-2}$\\
15.70&-8.04&1.93&$1.08\times10^9$&$2.30\times10^{-2}$&$\le 10^{-3}$& 2.23 &
$1.04\times10^{-2}$\\
15.48&-7.94&2.27&$1.06\times10^9$&$1.02\times10^{-2}$&$\le 10^{-3}$& 2.00 &
$4.64\times10^{-3}$\\
\noalign{\smallskip}\hline\noalign{\smallskip} 
\end{tabular}
\end{center} 
Notation as in Table 1.
\end{table*}    

\begin{figure}[ht]
\includegraphics[width=7.8cm]{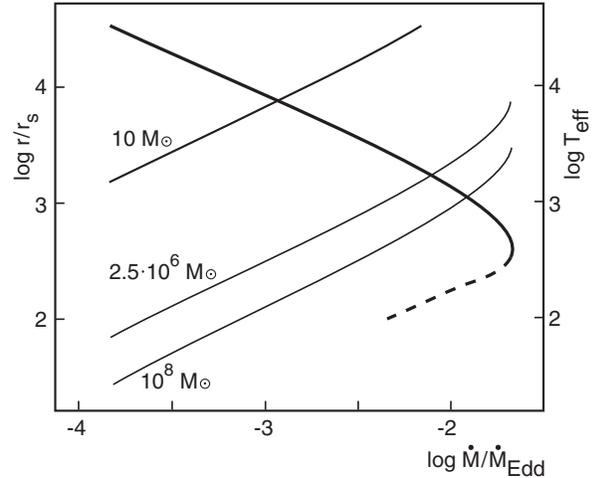}
\caption{ Solid line: radius of transition from accretion via a geometrically
thin disk to a coronal flow/ADAF (in units of Schwarzschild radius
$r_{\rm S}$) as function of the mass flow rate in the corona
(in units of the Eddington accretion rate $\dot M_{\rm {Edd}}$),
the dashed part of the relation is only an indication for a decreasing
evaporation efficiency (reliable results require the consideration of
a two-temperature corona). Thin lines:
$T_{\rm{eff}}$ effective temperature at the inner edge of the thin
disk as function of radius $r_{\rm S}$ and accretion rate $\dot M$, (see text)
.}
\end{figure}

In Fig. 1 we show the relation between the location of the inner disk
edge and the accretion rate. For example for the mass flow rate\
$\dot M/\dot M_{\rm {Edd}}$=$10^{-3}$ the thin disk is 
truncated at about 6$\cdot10^3$  Schwarzschild radii. Inside the
accretion flow is hot and optically thin, the thin disk has a
``hole''.

Also shown in Fig. 1 is a measure for the effective temperature
$ T_{\rm{eff}}^4= 3GM\dot M /{8 \pi \sigma r^3}$ near the inner edge of
the thin disk. The effective temperature does not scale. The higher
the black hole mass the lower is the
temperature. These disk temperatures are important for the spectral
fits. For comparison we also show the temperature in a disk around a
stellar black hole.

The recent investigations of R\'o\.za\'nska \& Czerny (2000a, 2000b) mostly
concentrate on the hot flow properties and use a two-temperature
plasma. They find a slightly different relation between evaporation
rate and distance. We investigate the equilibrium where the assumption
of a one-temperature plasma is a still valid.

\section {Comparison with observations}

The investigation of nuclei of bright nearby elliptical galaxies by Fabian
\& Canizares (1988) already showed that the galactic nuclei are very
under luminous. As suggested by Fabian \& Rees (1995)  and
confirmed by detailed calculations of Mahadevan (1997) the problem
is resolved if the galactic nuclei accrete via an ADAF. Likewise the apparently
contradictory observations of Sgr$\rm{A^*}$ could be understood by
advection-dominated accretion (suggestion by Rees (1982), first
spectral model by Narayan et al. (1995)). Further successful
applications of the ADAF model were carried out for the maser galaxy NGC 4258
and the nearby galactic nuclei M87 and M60 (for a review see Narayan et
al. 1998).

Recent observations in X-rays and in additional different wavelength bands
brought further information and at the same time new constraints.
Improvement also comes from new measurements of accurate mass and
distance, e.g. for NGC 4258 (Herrnstein et al. 1997, 1999).

Di Matteo et al. (1999) examined high-frequency radio observations
of three giant low-luminosity elliptical galaxies in the Virgo
cluster NGC 4649, NGC 4472 and NGC 4636 using the VLA (Very Large
Array) and SCUBA (Submillimetre
Common-User Bolometer Array) on the Clerk Maxwell Telescope, and hard
X-ray emission. They found radio fluxes lower than the emission predicted
by the standard ADAF model. In a second paper Di Matteo et al. (2000)
discuss constraints for the advection-dominated flow in the three Virgo
ellipticals and three cluster galaxies M87 (NGC 4486), NGC 1399, and
NGC 4696 and suggest that mass loss is important in the
radiatively inefficient accretion flow. 

In Fig. 2 we summarize the results to show the wide range
of values allowed from spectral fits.
In investigations including wind loss the
quantity $r_{\rm{out}}$ is used instead of a transition radius and
the wind loss is parametrisized by assuming that the mass flow rate
decreases as $\dot M=\dot M_{\rm{out}}\cdot (r/r_{\rm{out}})^p$.
For some galaxies
only fits for very different truncation radii were investigated,
e.g. radii of 300 or $10^4$ $r_{\rm S}$ were considered by Di Matteo
et al. (2000) for the elliptical galaxies, together with rates
$\dot M/\dot M_{\rm{Edd}}$ in the range 0.01 to 0.001. We show their
best and second best fit.
We also show the
results from recent ADAF modeling of $\rm{Sgr A^*}$ and the very
similar galactic transient source V404 Cyg (Quartaert \& Narayan 1999)
which explains the spectra well without wind loss. Also the
observations for M87 can be modeled without mass loss (Di Matteo et
al. 2000). Quataert et al. (1999) pointed out that for accretion rates
of 0.03 $\dot M_{\rm{Edd}}$ the thin accretion disks in M87 and NGC 4696
must be truncated at $\ge 10^4$ Schwarzschild radii.

For $\rm{Sgr A^*}$, M87 (fit without wind loss) and NGC 4649
predictions and fit results agree reasonably.
For elliptical galaxies the spectral fits were performed only for
radii much larger or much smaller than the transition radius predicted
from our model, which makes the comparison difficult. Quataert
\& Narayan (1999) argued that infall of gas from a cooling flow could
change the location of the truncation. Such a boundary
condition is not included in our analysis.
But new observations by Chandra might anyhow
change the picture. Mushotzky et al. (2000) report on observations of 4 giant
elliptical galaxies, NGC1399, NGC1401, NGC4472 and NGC4636 and the
possibility to set upper limits for the radiation, but quantitative
results are not yet presented.

For the low luminosity AGN (LLAGN) M81 and NGC 4579 the observed UV flux 
demands a truncation  far inward, which is in contradiction to our
findings. Quataert et al. (1999) already pointed out that it needs to
be explained how for similar mass flow rates such different transition
radii are found, about 100 $r_{\rm {S}}$ for the LLAGN and $10^3$ to $10^4 
r_{\rm {S}}$ for nuclei of elliptical galaxies. These authors raised the
question whether
the cooling flow in ellipticals could be responsible for this
difference. But comparing with our results the LLAGN are the systems
which need to be explained. Possibly the evaporation picture can
offer an alternative view as will be discussed in the next section.
For a review of UV and optical continuum emission in
AGN see Koratkar \& Blaes (1999).

\begin{figure}[ht]
\includegraphics[width=8.8cm]{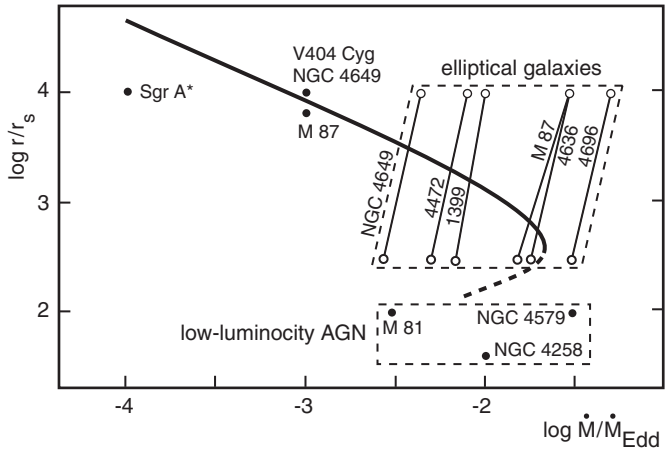}
\caption{Comparison of the theoretical $r$-$\dot M$ relation (same as
in Fig. 1) with results from fits of observed spectra. Mass accretion
rates and radii from various investigations:
elliptical galaxies from Di Matteo et al. (2000) (best and second-best
fit), low-luminosity AGN M81 and NGC 4579 from Quataert et
al. (1999), NGC 4258 from Gammie et al. (1999), Sgr $\rm{A^*}$,
NGC 4649 and the same value for V404 Cyg from Quataert \& Narayan
(1999), value for M87 (without wind) from Reynolds et al. (1996).
Filled symbols correspond to fits without wind loss, open symbols
with wind loss.} 
\end{figure}

\section {Discussion and conclusions}
We discussed how the transition radii predicted by our model 
compare to the results for the truncation found from the spectral fits
using an ADAF for the innermost region and a geometrically
thin disk further out. We find three different groups: (1) agreement 
with our prediction is found for M87 (model without wind loss)
and NGC 4649 (see also Liu et al. 1999). We want to point out the
good agreement of our
$r$-$\dot M$ relation for the stellar black hole source V404
Cyg, the soft X-ray transient with the best X-ray data available.
The agreement is generally good in the case of
X-ray novae (Meyer et al. 2000b). This is important since the
constraints are more severe for the
galactic black hole binaries. The observed $\rm{H\alpha}$ emission lines
give the maximum rotational velocity in the disk and therefore
indicates the truncation. (2) The comparison for nuclei of
elliptical galaxies is difficult since the radii taken for spectral
fits are much larger or smaller than the radii expected from our predictions.
A better comparison will be possible when the new
Chandra observations (Mushotzky et al. 2000)
will become available. (3) A clear discrepancy is found for the low
luminosity AGN (LLAGN), especially M81.

Concerning the discrepancy found for LLAGN we discuss in the
following whether one could imagine other sources of UV radiation
than a geometrically thin standard disk. In the modeling
the evaporation efficiency has a maximum around a critical radius of
300 $r_{\rm S}$ (the number depends on the chosen value of the
viscosity in the hot coronal gas, $\alpha$=0.3, and might be
influenced by simplifications in the computational method). If the
mass accretion rate is sufficiently low
the thin disk should be truncated there or further outward.
The decreasing evaporation efficiency inside the critical radius
suggests that inside mass could settle again in a thin disk in the
equatorial plain.
We have not yet included the physics of a two-temperature gas which
becomes important with decreasing distance from the black hole.
R\'o\.za\'nska \& Czerny (2000a) discuss the condensation of matter
from the hot coronal flow in a cool disk below. The possible existence
of such an interior disk was also mentioned in connection with
spectral transitions (Meyer et al. 2000a). We note that such a
disk could be cool and the mass flow rate within this disk could be low 
(an ``inert'' disk), so that most of the mass flow remains in the
ADAF. Spectral observations of a reflection component and the Fe
$\rm{K}{\alpha}$ line, observed for many black hole sources (see e.g.
Gilfanov et al. 1999), led to detailed discussions of the accretion
flow geometry and give evidence for an inner disk below the ADAF region. 
In this context it is interesting that Vilhu et al. (2000) modeled
observations for the galactic black hole source GRS 1915+105 with a
corona above an interior disk.
The interaction of hot protons in the ADAF with such an inner disk
underneath could transfer accretion energy from the ADAF to the disk
where it is radiated away in various wavelength bands, 
depending on the black hole mass (compare the effective temperatures
at the transition radius, Fig. 1). The observed UV radiation then
would not necessarily point to a disk truncation at 100 Schwarzschild
radii.

\begin{acknowledgements}
We thank Friedrich Meyer for discussions.
\end{acknowledgements}

\end{document}